\documentclass[doublecol,cite]{epl2} 

\usepackage{graphicx,epsfig}
\usepackage{subfigure}  
\usepackage{amsmath}
\usepackage{amsfonts}
\usepackage{amssymb}
\usepackage{xfrac}
\usepackage{enumitem}
\usepackage{xcolor}
\usepackage{color}
\usepackage{epstopdf}
\usepackage{ulem}
\usepackage[hidelinks]{hyperref}
\usepackage[capitalise]{cleveref}
\usepackage{bbold}
\usepackage{fancyhdr}

\newcommand{\Da}{\mathcal{D}}

\title{Chiral Active Matter}
\date{\today}

\shorttitle{Chiral Active Matter} 

\author{Benno Liebchen\inst{1} \and Demian Levis \inst{2,3}}
\shortauthor{B. Liebchen, D. Levis}

\institute{                    
  \inst{1} Institute of Condensed Matter Physics, Technische Universität Darmstadt -
D-64289 Darmstadt, Germany\\
  \inst{2} Departament de F\'isica de la Mat\`eria Condensada, Universitat de Barcelona - C. Mart\'i Franqu\`es 1, 08028 Barcelona, Spain.\\
  \inst{3} UBICS University of Barcelona Institute of Complex Systems - C. Mart\'i i Franqu\`es 1, E08028 Barcelona, Spain.
}


\abstract{
Chiral active matter comprises particles which can self-propel and self-rotate. 
Examples range from sperm cells and bacteria near walls to asymmetric colloids and pea-shaped Quincke rollers.
In this perspective article we focus on recent developments in chiral active matter.
After briefly discussing chiral active motion at a single particle level, we
discuss 
collective phenomena 
ranging 
from vortex arrays and patterns made of rotating micro-flocks to states featuring unusual rheological properties. 
\vskip 0.1cm 
\textbf{EPL perspective article invited by A. Lanotte}
\vskip -0.7cm
}

\begin{document}
\maketitle

\section{Introduction}
In his speech about the molecular tactics of a crystal in 1893, Lord Kelvin defined an object as chiral if its image in a plane mirror cannot be brought to coincide with itself. 
When using corkscrews or scissors, wearing shoes or observing snail houses, we frequently encounter chirality in our everyday live -- and it fact, when looking at our hands we see a chiral object and its mirror image at the same time. Accordingly, we commonly refer to 
"left-" and "right-handed" representatives of a chiral object. 
\\Chirality can not only occur in the shape of an object, but also in its motion or trajectory. 
Such chiral motion has been scientifically described and analyzed by 
Bronn \cite{bronn1862klassen} and Jennings \cite{jennings1901significance} more than a century ago. It now receives a renewed interest
~\cite{teeffelen2008dynamics,Kummel2013Circular,lowen2016chirality,Liebchen2017Collective,levis2019activity,levis2019simultaneous,denk2016active,liao2018clustering,huang2020dynamical,kruk2020traveling,ventejou2021susceptibility,Chen2017Weak,Kim2018Large,oliver2018synchronization,Hernandez2020Collective, Hokmabad2022Spontaneously,afroze2021monopolar,heckel2020,markovich2019chiral,ma2022dynamical,fruchart2021non,o2017oscillators,hoell2017dynamical,ai2018mixing,maitra2019spontaneous,maitra2020chiral,lei2019nonequilibrium,wang2021emergent,liao2018transport,reichhardt2019reversibility,lowen2019active,liu2019configuration,o2019review,liu2019collective,kole2021layered,supekar2021learning,beppu2021edge,moore2021chiral,kruk2021finite,zhang2021persistence,reigh2020active,hernandez2022dynamics,mijalkov2013sorting,kurzthaler2017intermediate,chepizhko2020random,scholz2021surfactants}
in the research field of active matter \cite{marchetti2013hydrodynamics,bechinger2016active,chate2020dry,hecht2021introduction,liebchen2021interactions}, which describes the motion and collective behaviour of self-propelled agents, such as 
bacteria, algae or synthetic Janus colloids.
\\\underline{Linear and chiral active matter:}
In this article we distinguish three classes of active particles. Members of the first class self-propel linearly and change their direction of motion randomly, often due to thermal fluctuations, and in an unbiased way. The motion of individual linear active particles is largely characterized by their self-propulsion speed relative to the strength of the fluctuations in their environment. 
The second class comprises particles (or agents) which can both self-propel and self-rotate. They feature a 
characteristic self-propulsion velocity and a characteristic frequency 
determining the rate of change of their direction of motion in the absence of fluctuations. This leads to trajectories which are chiral in the sense that their mirror images occur with a different (lower) probability in a representative particle ensemble, even in bulk. 
In the absence of fluctuations, chiral active particles (CAPs) typically move along circles (2D) or helices (3D).
Members of this class comprise 
helically swimming sperm cells \cite{riedel2005self, friedrich2007chemotaxis}, malaria parasites \cite{patra2022collective},  chiral microtubules driven by molecular motors \cite{afroze2021monopolar}, bacteria which swim in circles near walls and interfaces ~\cite{diluzio2005escherichia,lauga2006swimming,Leonardo2011Swimming}, 
shape-asymmetric colloidal microswimmers~\cite{Kummel2013Circular,ten2014gravitaxis,Shelke2019Transition,Zhang2020Reconf,alvarez2021reconfigurable} and granular ellipsoids \cite{barois2020sorting,arora2021emergent},
as well as certain 
motile droplets \cite{Kruger2016Curling, Lancia2019Reorientation,Wang2021Active}
that (spontaneously) break chiral symmetry. 
Finally, the third class of active particles comprises so-called "spinners" that rotate but do not self-propel; i.e. they are essentially spinning tops that only move due to thermal (Brownian) motion, external forces and interactions with other particles. When coupling rotational and translational degrees of freedom, the latter can make spinners active at the many particle level \cite{van2016spatiotemporal,scholz2018rotating,workamp2018symmetry,soni2019,massana2021arrested,bililign2022motile}. 
\\\underline{How does chiral active motion emerge?}
In general, besides a continuous energy source, chiral self-propulsion needs a twofold symmetry breaking: one to break the fore-aft symmetry in order to create self-propulsion and one breaking the symmetry with respect to the self-propulsion direction. 
\begin{figure*}[h]
    \centering
    \includegraphics[width=1.\textwidth]{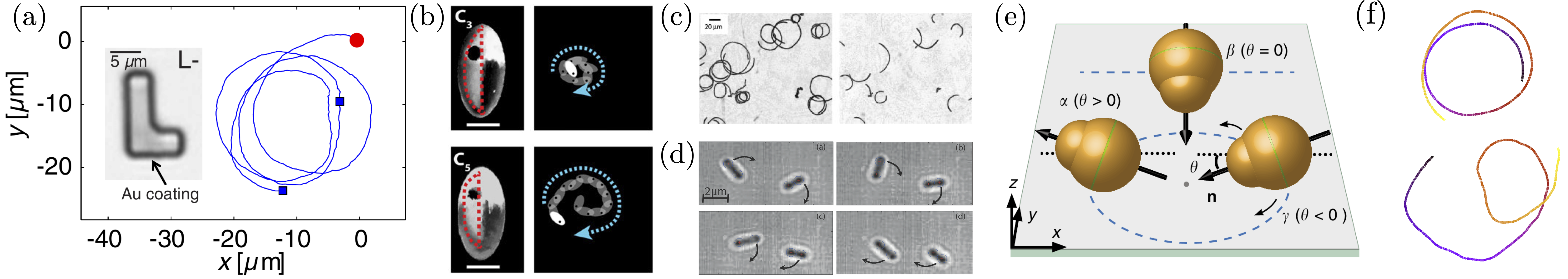}
      \caption{\small Single particle trajectories of (a) L-shaped colloidal  swimmers (adapted from \cite{Kummel2013Circular}), (b) chiral self-propelled granular ellipsoids (adapted from  \cite{arora2021emergent}) (c,d) E. coli bacteria  swimming near a glass surface (adapted from \cite{lauga2006swimming} and \cite{oliver2018synchronization}, respectively), (e) 
      schematic illustration of pear-shaped Quincke rollers
 (adapted from \cite{Zhang2020Reconf}) 
      and (f) simulated single CAP (following eq. \ref{CAP2}), with $l_P/R=100 $ and  $\Omega=100$ (top), and $l_P/R=10$ and $\Omega=10$ (bottom)  
       (color code indicates time evolution, black being the initial position).
}
    \label{fig:chiralABPtrajs}
\end{figure*}
Accordingly, the most obvious way towards chiral self-propulion is to use active particles with chiral shapes (or with chiral surface properties). Such a shape couples the translational and rotational degrees of freedom of a particle \cite{kraft2013brownian} and leads to chiral trajectories, as illustrated by the  L-shaped colloidal microswimmers \cite{Kummel2013Circular} in Fig. 1.
A variant of this route is based on using achiral (symmetric) particles with anisotropic interactions allowing the particles to self-assemble into structures with chiral shapes
\cite{ebbens2010self,nourhani2016spiral,wykes2016dynamic,vutukuri2017rational,
aubret2018targeted,aubret2018diffusiophoretic, alvarez2021reconfigurable}. Similarly, also binary mixtures of isotropic particles with non-reciprocal interactions can self-assemble into chiral clusters showing chiral active motion
\cite{schmidt2019light,grauer2021active}.
Other mechanisms leading to chiral self-propulsion 
hinge on hydrodynamic interactions of linear swimmers with walls or interfaces, as illustrated by circularly swimming \emph{E.coli} bacteria 
\cite{lauga2006swimming} or exploit memory effects in viscoleastic environents \cite{narinder2018memory}.
Finally, chiral active motion can also be achieved for isotropically shaped droplets in simple environments, based on the integration of 
chiral molecular motors \cite{Lancia2019Reorientation} or even spontaneously by exploiting phase separation
\cite{Wang2021Active}.
\\\underline{Content:}
In this perspective article, we discuss recent progress in the realization, modeling and understanding of chiral active matter. We'll largely focus on the collective behavior of CAPs, and comment only briefly on 
the dynamics of individual particles  (see the earlier review \cite{lowen2016chirality} for a comparatively detailed discussion) and  spinners systems. 

\section{Single CAP}
The simplest model to describe chiral active Brownian motion of an individual circle swimmer in two dimensions (2D), 
with position $\vec r(t)
$ and orientation $\vec p(t) = (\cos\theta(t),\sin\theta(t))$, is based on the following overdamped Langevin equations: 
\begin{equation}
\dot{ \vec{r}}=v_{0} \vec{p}+\sqrt{2 D} \vec{\xi} 
;\qquad
\dot{ \theta}=\omega+\sqrt{2 D_{\mathrm{R}}} \eta \label{CAP2}
\end{equation}
Here $v_0$ and $\omega$ are the characteristic self-propulsion speed and angular velocity (or  frequency) of the swimmer. 
The second term on the right hand side of both equations represents Brownian motion, where 
$\eta(t)$ and the components of $\vec \xi(t)=\left(\xi_x(t),\xi_y(t)\right)$ are Gaussian white noise variables with zero mean and unit variance and $D$ and $D_{\mathrm{R}}$ 
are the translational and rotational diffusion coefficients, respectively.  
Choosing length and timescales as $\tau_{\mathrm{p}}=D_{\mathrm{R}}^{-1}$ and $l_{\mathrm{P}}=v_{0} D_{\mathrm{R}}^{-1}$ and defining $\vec r^\ast=\vec r/l_{\mathrm{P}}$ and $(\vec \xi^\ast,\eta^\ast)=(\vec \xi,\eta)/\sqrt{\tau_{\mathrm{p}}}$
yields, after omitting the asterixes
$\dot{\vec{r}}=\vec{p}({t})+\mathrm{Pe}^{-1} \vec{\xi}; \quad \label{eq:1CAPn}
\dot{\theta}={\Omega}+\sqrt{2} \eta$. 
Thus, chiral active Brownian motion of a single particle depends on two dimensionless parameters only: (i) the P\'eclet number $\mathrm{Pe}=\frac{v_{0}}{\sqrt{2 D D_{\mathrm{R}}}}$ (where $\mathrm{Pe} \propto \frac{l_{\mathrm{P}}}{R}$ 
measures the persistence length $l_{\mathrm{P}}$ of a swimmer (for $\omega=0$) in units of the particle radius $R$, when assuming Stokes-Einstein relations)
and
(ii) $\Omega = \omega/D_{\mathrm{R}}$ which compares the strength of deterministic and stochastic contributions to the angular velocity.
\\Exemplaric trajectories as obtained from these equations in Brownian dynamics simulations are shown
in Fig.~\ref{fig:chiralABPtrajs}f.
The motion of these swimmers can also be characterized analytically in terms of the mean trajectory, which takes the form of a 
spira mirabilis (logarithmic spiral) in 2D \cite{Kummel2013Circular,lowen2016chirality} and of a 
concho spiral in 3D \cite{wittkowski2012self}.

\section{Isotropic interactions} 
Isotropic interactions between CAPs can be easily accounted for by introducing an interaction potential $U$ which only 
affects the center of mass motion of the particles, such that Eqs.~(\ref{CAP2}) change to  
\begin{equation}\label{eq:CAPiso1}
\dot {\vec{r}}_i=v_{0} \vec{p}_{i}-\frac{1}{\gamma} \nabla_{\vec{r}_{i}} U +\sqrt{2 D} \vec{\xi}_{i}, \quad
\dot{\theta}_{i}=\omega+\sqrt{2 D_{\mathrm{R}}} \eta_{i}.
\end{equation}
Here $i=1...N$ labels particles, $\gamma$ is the damping (drag) coefficient and 
the interactions are typically taken to be isotropic, short-ranged and repulsive; e.g. in the form of Weeks-Chandler-Anderson interactions \cite{weeks1971role}. Such a volume exclusion introduces a new control parameter, the packing fraction $\phi=\pi\sigma^2 N/(4L^2)$, where $\sigma$ and $L$ refer to the linear size (diameter) of the particles and the system, respectively.
\\When $\omega=0$, this model reduces to the very-well explored model of 
linear (or achiral) active Brownian particles (ABPs), which show Motility-Induced Phase Separation (MIPS) at large enough Pe and $\phi$ \cite{CatesTailleurRev}. That is, remarkably, despite featuring purely repulsive interactions, active particles can phase-separate into a dense region and a coexisting gas.
\\A question that has attracted a significant interest in the past few years is how MIPS is affected by chirality (i.e. $\omega\neq0$). Both, studies based on particle-based simulations  \cite{liao2018clustering,mani2022} and on the analysis of continuum equations  \cite{bickmann2020analytical,ElenaJCP} found that circular swimming (in 2D) tends to hinder MIPS. 
Remarkably, however, chirality also leads to a new phenomenon which does not occur for $\omega=0$: the emergence of finite dynamical clusters which counter-rotate with respect to the surrounding gas \cite{liao2018clustering,mani2022}. 
(Notably, similar rotating finite clusters have also been observed in motile bacteria with depletion interactions \cite{schwarz2012phase}.)
\\By explicitly coarse-graining the microscopic dynamics, eqs. (\ref{eq:CAPiso1}), and performing a mean-field-like closure scheme \cite{BialkeEPL,bickmann2020analytical,mani2022,ElenaJCP}, the following equations can be derived for a coarse-grained version of the microscopic density $\rho(\vec{r},t)=\sum_i \delta(\vec{r}-\vec{r}_i)$ and the polarization density $ \vec{w} (\vec{r},t)=\sum_i \delta(\vec{r}-\vec{r}_i)\vec{p}_i(t)$:
\begin{equation}
\begin{split}
 \dot \rho =- \nabla \cdot \Big[ v_{\bar{\rho}} \vec{w} -\Da \nabla \rho \Big]
\end{split}
\label{modeldescrip_eq:15}
\end{equation}
\begin{equation}
\begin{split}
 \dot {\vec{w}} &= -\nabla \cdot \Big[\frac{\mathbb{1}}{2} \rho v_{\bar{\rho} } - \Da \nabla \vec{w} \Big]  - (\omega\mathcal{R}  + D_{r}) \vec{w} 
 \end{split}
\label{modeldescrip_eq:16}
\end{equation}
where  
$\mathcal{R}$ corresponds to a $\pi/2$-rotation and $\Da$ is an effective many-body diffusion coefficient. The effective velocity $v_{\bar{\rho}} = v_0 -\bar{\rho} \zeta$ decays linearly with density with an effective "friction" parameter $\zeta\geq0$. This is the key aspect of the theory leading to particle aggregation and eventually MIPS: particles block each other in the direction of self-propulsion, giving rise to a reduction of $v_{\bar{\rho}}$. 
The linear stability analysis of the homogeneous disordered solution of these field equations predicts a spinodal-like, long-wavelength instability \cite{BialkeEPL,bickmann2020analytical,mani2022,ElenaJCP}, and a  short-wavelength one \cite{ElenaJCP}, interpreted as the onset of MIPS in the first case, and the formation of structures with a characteristic (onset) length scale $\ell$ in the second case \cite{ElenaJCP}. The spinodal-like instability shifts to higher values of $v_0$ (and $\zeta$) as $\omega$ increases, showing how chirality opposes MIPS. Notably, for the short-wavelength instability, the characteristic length scales as $\ell\propto v_0$, and, for large $\omega$ also as $\ell\propto 1/\omega$ \cite{ElenaJCP}. Interestingly, this linear increase of the onset cluster size with the single-particle swimming radius ($\ell \propto v_0/\omega$)
is the same scaling law which occurs for CAPs with alignment interactions as we'll see below.
\\Another remarkable phenomenon which can occur for CAPs with isotropic interactions is a chirality-induced absorbing-active transition towards a hyper-uniform state, characterised by vanishing long-wavelength density fluctuations \cite{lei2019nonequilibrium}.

\begin{figure}
    \centering
    \includegraphics[width=.48\textwidth]{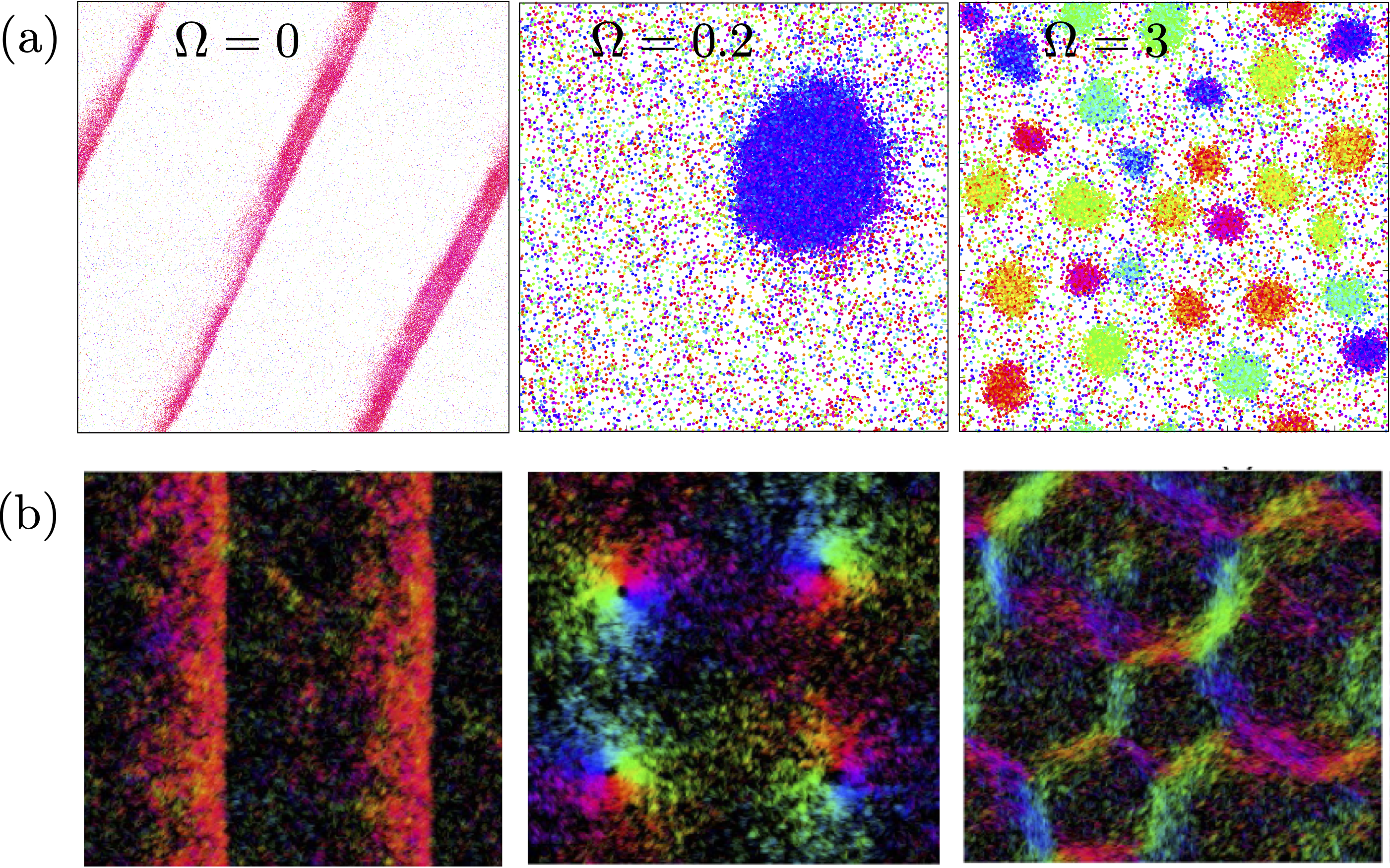}
    \caption{\small Simulation snapshots of CAPs with colors encoding particle orientations:
    (a) Particles with polar alignment and a single angular frequency (Eqs.~(\ref{cap})) can self-organize into traveling bands $(\Omega=0)$, rotating macroflocks $(\Omega=0.2)$ and micro-flock patterns $(\Omega=3)$. 
     (b) Particles with a symmetric distribution of angular frequencies \cite{ventejou2021susceptibility} can form
     (from left to right): traveling bands (surviving weak frequency dispersion) and polar vortices (for polar alignment) as well as active foams (for nematic alignment).
     Figures adapted from refs. \cite{Liebchen2017Collective} (a) and \cite{ventejou2021susceptibility} (b).  
    }
    \label{fig:cap}
\end{figure}

\section{Polar alignment interactions}
A minimal
model describing aligning CAPs, originally called the "CAP model", has been introduced in ref.~\cite{Liebchen2017Collective}. It can be viewed as a chiral generalization of 
the paradigmatic Vicsek model \cite{Vicsek1995} with additive interactions \cite{farrell2012pattern,martin2018collective,chepizhko2021revisiting} and in continuous time:
 \begin{equation}
\dot{\vec{r}}_{i}=v_0 \vec{p}_{i}; \quad 
\dot{\theta}_{i}=\omega+\frac{K}{\pi R^2} \sum_{j \in \partial_{i}} \sin \left(\theta_{j}-\theta_{i}\right)+\sqrt{2 D_{\mathrm{R}}} \eta_{i}
\label{cap}
\end{equation}
Here $K$ controls the interaction strength and the sum is performed over all particles with a distance up to $R$ (interaction range) from the $i$-th particle\footnote{Here the interaction term is normalized by the interaction area. Variants of the CAP model such as the one explored in ref.~\cite{ventejou2021susceptibility}, use instead the number of neighbours to normalize the interactions.}. 
The model has four independent parameters, which are given by the following dimensionless parameters: 
the (reduced) packing fraction $\rho_{0}=N R^{2} / L^{2}$, the rotational P\'eclet number $\mathrm{Pe}_{\mathrm{r}}=v_0 /\left(D_{R} R \right)=l_{\mathrm{p}}/R$ 
and
$g=K /\left(\pi R_{\theta}^{2} D_{R}\right)$, $\Omega=\omega / D_{R}$, which compare the alignment rate and the intrinsic reorientation rate with rotational diffusion.
Remarkably, however, the phenomenology of the CAP model is largely determined by two effective parameters, the combined parameter $g\rho_0$ and $\Omega$, whereas $\mathrm{Pe}_{\mathrm{r}}$ mainly manifests in the length scales of the emerging structures. 
\\For $\Omega=0$ the CAP model reproduces the phenomenology of the Vicsek model; i.e. for $g\rho_0 \lesssim 2$ one observes a disordered uniform state and for $g\rho_0 \gtrsim 2$ high-density bands featuring polar order and traveling through a disordered low-density background (Fig. ~\ref{fig:cap}a). Finally, $g\rho_0\gg 2$ leads to a Toner-Tu-like phase, showing global polar order and very strong density fluctuations. 
\\\underline{Macro-droplets and micro-flock patterns:} For finite $\Omega \ll 1$, instead of bands, we observe rotating macro-droplets 
within which most of the particles phase-synchronize (see Fig.~\ref{fig:cap}b where colors show the momentaneous orientation of the particles). 
The most striking feature of the CAP model is that it leads, for larger $\Omega$ ($\Omega \gtrsim 1$), to pattern formation. The pattern comprises high-density clusters made of phase-synchronized CAPs featuring a characteristic length scale (Fig.~\ref{fig:cap}c) which is invariant to changes of the system size (at least at early and intermediate times). 
That is, for small $\Omega$ and $g\rho_0>2$ the system selects a characteristic density while, for large $\Omega$, it selects a length scale (the size of the micro-flocks).
Note that similar rotating clusters have also been observed in experiments \cite{riedel2005self, loose2014bacterial, patra2022collective} (see Fig.~\ref{fig:sperm}) and in
simulations of curved active polymers  \cite{denk2016active}. 
\\The described phenomena can be broadly understood based on continuum theories of the CAP model and similar ones \cite{Liebchen2017Collective,kruk2020traveling, ventejou2021susceptibility}. Such theories can be derived 
based on systematic coarse graining e.g. via Dean's formalism \cite{dean1996langevin}, or Fokker-Planck equations \cite{risken1996fokker}, which after moment expansions 
supplemented with suitable closure schemes 
\cite{bertin2006boltzmann,peshkov2014}, lead to a closed system of equations for the particle density $\rho(\vec r,t)$ and the polarization density $\vec w(\vec r,t)$
field. For the CAP model \cite{Liebchen2017Collective}, the lowest order terms read (here written in dimensionless form to simplify the interpretation):
\begin{equation}
    \dot{\rho}= -\mathrm{Pe}_{\mathrm{r}}\,\nabla \cdot {\vec w} \label{rho1}
\end{equation}
\begin{equation}
\dot {\vec w} = \left(g \rho-2\right)\frac{{\vec w}}{2} + \Omega {\vec w}_\perp - \frac{\mathrm{Pe}_{\mathrm{r}}}{2}\nabla \rho 
+ \text{h.O.} 
\label{pold}
\end{equation}
The first term on the right hand side of Eq.~(\ref{pold}) shows that the uniform disordered solution $(\rho,\vec w)=(\rho_0,\vec 0)$ (base state) looses stability at $g\rho_0=2$, leading to polarized states.  
The second term comprises the subscript $\perp$ which represents a $\pi/2$-rotation and
causes a uniform rotation of aligned particles (e.g. within the rotating macro-droplets or the micro-flocks). 
Finally, 
in the third term, and all neglected higher-order terms (h.O.), the $\nabla$ operator and the rotational P\'eclet number show up as (powers of) the product $\mathrm{Pe}_{\mathrm{r}} \nabla$ only. That is, after transforming (the linearzed equations) to Fourier-space ($\nabla \to - \operatorname{i}\vec q$, where $\operatorname{i}$ is the imaginary unit), $\mathrm{Pe}_{\mathrm{r}}$ can be absorbed in the wavevector $\vec q$ ($\vec {\tilde q}:=\mathrm{Pe}_{\mathrm{r}} \vec q$) such that it does not affect the linear stability criteria but leads to a generic scaling of the characteristic onset length scale $\ell$ of the emerging structures with $\ell \propto \mathrm{Pe}_{\mathrm{r}}$.
A more detailed analysis shows that $\ell$ also scales with $1/\Omega$, ultimately leading to the scaling $l \propto \mathrm{Pe}_{\mathrm{r}}/\Omega \propto v_0/\omega$ which matches with observations of the micro-flock-scaling in particle based simulations \cite{Liebchen2017Collective}.
\\\underline{Vortex and cloud states:} Notably, the authors of ref.~\cite{ventejou2021susceptibility}
have performed a detailed analysis of a variant of the CAP model, which uses a uniform noise and normalizes the  interaction term by the local density. In particular, this work has identified an additional vortex phase. A similar vortex phase has also been seen in other variants of the CAP model, such as ref.~\cite{kruk2020traveling} which has also observed additional cloud-like structures, 
and in ref.~\cite{liao2021emergent} where CAPs with dipolar interactions have been explored.
\\\underline{Other interactions:} The CAP model (Eqs. \ref{cap}) has also been extended to account for short-ranged repulsions \cite{levis2018micro}. In this case micro-flock pattern formation survives, 
but the resulting rotating clusters are strongly anisotropic. 
Several works have explored the role of other (non-polar or ferromagnetic-like) alignment interactions, such as nematic \cite{ventejou2021susceptibility}, phoretic interactions \cite{liebchen2016pattern} and dipolar interactions \cite{liao2021emergent}.
For nematic interactions, in particular, nematic vortices and "active foams" have been observed (see Fig. \ref{fig:cap}b).
Comparatively complex interactions have been explored also experimentally 
in ensembles of 
pear-shaped Quincke rollers \cite{Zhang2020Reconf}
which move in circles on a substrate when an external electric field is applied (see Fig. \ref{fig:chiralABPtrajs}e). These rollers show various phases such as a disordered gas phase, rotating flocks and also vortices where the particles have aster-like orientations.
\\\underline{Mixtures and synchronization:} 
The CAP model, eqs.~\ref{cap}, has also been generalized to account for different angular frequencies  
\cite{levis2019simultaneous,ventejou2021susceptibility} 
which leads to intriguing combinations of phase separation and pattern formation \cite{liebchen2016pattern}. In particular, binary mixtures of CAPs of opposite handedness tend to segregate and can show, for example, micro-flock pattern formation in one species and macro-droplet formation in the other species, such that the system simultaneously selects a characteristic density and characteristic length scale \cite{levis2019simultaneous}. 
Notably, segregation or self-sorting has also been observed in simulations and experiments of elongated CAPs \cite{arora2021emergent}, in which alignment interaction emerge from steric effects  \cite{liu2019collective,arora2021emergent}.
\\An important problem for mixtures of CAPs is to understand under which conditions they synchronize their frequencies. 
In ref.~\cite{levis2019activity} CAPs have been considered as motile phase oscillators and it has been shown that activity can support, or even induce, synchronization in parameter regimes where non-moving oscillators (or oscillators moving by passive diffusion) would not synchronize. In particular, the work has identified a 
 positive feedback loop involving a two-way coupling between the oscillators' phase and self-propulsion, which helps the particles to synchronize (in particular when the interaction among the particles is normalized by area and not by the local particle density). This can manifest in two different forms~\cite{levis2019activity}: 
 (i) The first one is best illustrated by binary mixtures of two counter-rotating species, which self-segregate and form oppositely rotating macro-droplets. Within each of the clusters, most of the particles share the same frequency and phase. The macro-droplets also feature a characteristic density and a size which scales with system size (asymptotically for late times), so that the particles within each macro-droplet synchronize over larger and larger distances when the system size is increased. Notably, for more general mixtures made of CAPs with a continuous frequency distribution, a similar self-segregation is observed; here, interestingly, the particles can even synchronize their frequencies within the macro-droplets. 
 (ii) The second form of synchronization occurs at large density and relatively low angular frequency. In this parameter regime, two species of opposite handedness can form the "mutual flocking" state, where the 
two species mutually suppress their rotations and move at a characteristic angle to each other. The resulting phase shows finite global polar order and can be viewed in some sense as a generalization of the Toner-Tu phase. 
Notably, 
ref.~\cite{ventejou2021susceptibility} reports on a numerically performed linear stability analysis predicting that in sufficiently large systems the mutual flocking phase is generally linearly unstable. Accordingly, to observe 
activity-induced synchronization one should probably focus on finite systems of sufficiently large density or on rotating macro-droplets. Understanding the precise criteria for which activity can induce synchronization presents an interesting topic for future explorations, to which we'll come back at the end of the article. 

\section{Hydrodynamic interactions}
\begin{figure}
    \centering
    \includegraphics[width=0.36\textwidth]{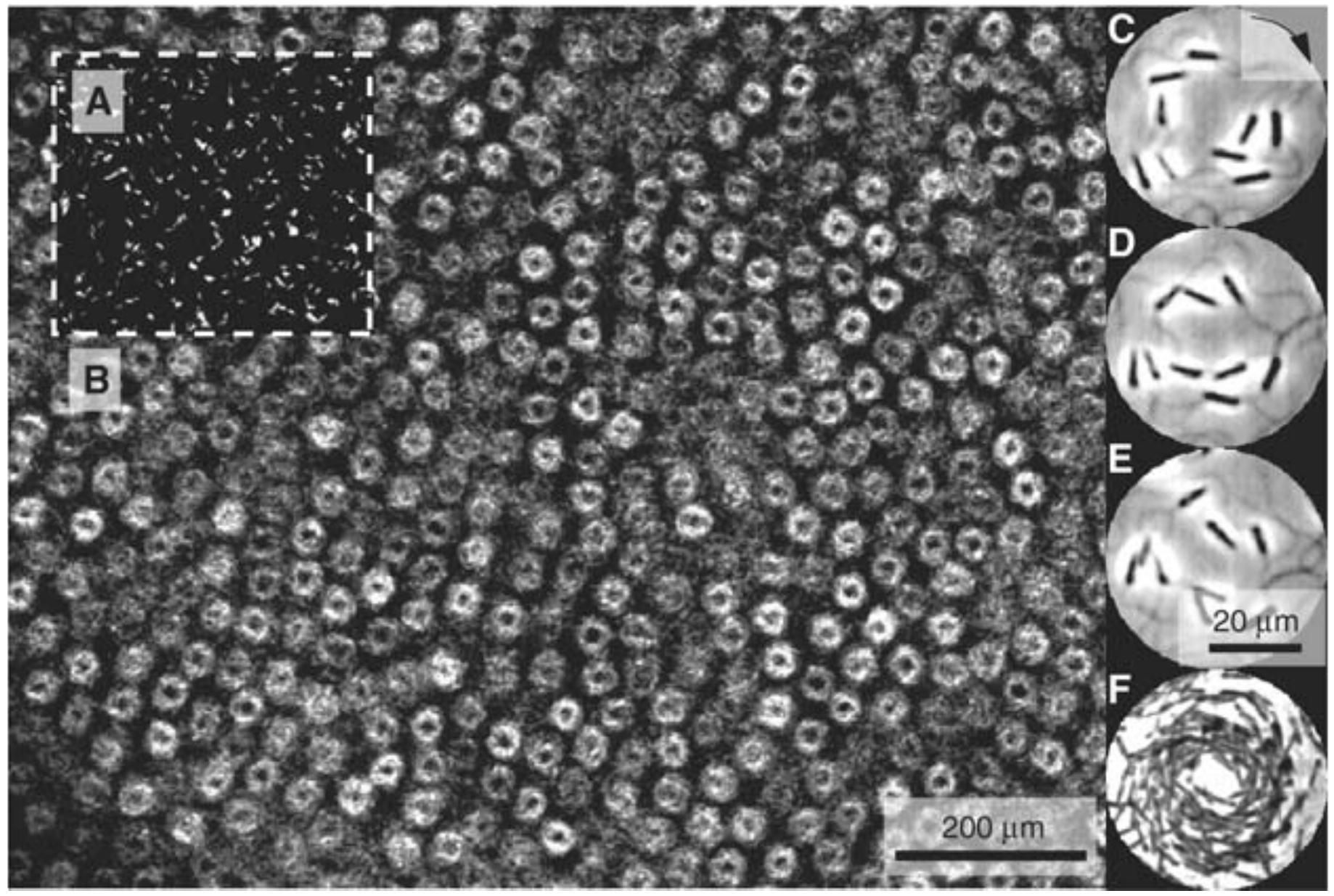}
    \caption{\small Dark field-contrast image of 
    circulating spermatozoa (A) and average over 25 consecutive frames showing a 2D array of rotating clusters (B). (C--E) Successive frames of a phase-contrast movie showing nine spermatozoa swimming clockwise (arrow). (F) Average of 25 frames. See \cite{riedel2005self} for details.
    }
    \label{fig:sperm}
\end{figure}

Many CAPs, from ciliated microorganisms to asymmetric colloids, are commonly found swimming in a fluid. 
As they move, they actuate the fluid around,
leading to 
hydrodynamic interactions (HI) among them. One  way of consistently modeling HIs, is to consider particles of spherical shape generating anisotropic surface flow velocities that make them self-propel. This squirmer model \cite{blake1971spherical},
which is commonly formulated with axisymmetric surface flows leading to linear swimming, has been recently extended to include swirling \cite{pedley2016squirmers}  and, more generally, non-axisymmetric surface flows leading to chiral propulsion (or spinning) \cite{paklauga,burada2022hydrodynamics,maity2022near}.
\\While the impact of 2D confinement, which leads to HIs between CAPs and walls or interfaces, has been the focus of several works \cite{lauga2006swimming, lauga2019transition, elgeti2010hydrodynamics},  studies on the collective behaviour of chiral squirmers with hydrodynamic cross-interactions 
are still rather scarce.
Examples include studies of 
small populations of chiral swimmers, showing `dancing states' in Volvox \cite{drescher2009dancing}, and synchronized  motion in rotating bacteria \cite{oliver2018synchronization}. 
Another study which explores the collective behaviour of CAPs with HIs based on 
lattice Boltzmann simulations shows the self-organization of 2D chiral swimmers in a variety of structures, ranging from small spinning  clusters (or molecules), to collectively moving larger structures \cite{shen2019hydrodynamic}.
An interesting experiment with CAPs where hydrodynamic cross-interactions play a key role has been discussed 
in ref.~\cite{petroff2015fast}. This work has
explored spherical 
{\it Thiovulum majus} bacteria (5-20$\mu m$ in size) which are covered with hundreds of tiny flagella and swim at speeds up to 600$\mu m/s$. The bacteria have been observed to self-organize into a 2D crystal of rotating cells \cite{petroff2015fast} resembling previous observations in 
sperm cells \cite{riedel2005self} (see Fig. \ref{fig:sperm}). 
The finding in ref.~\cite{petroff2015fast} has been rationalized based on the fact that the rotating bacterial flagella create a tornado-like flow that 
pulls neighboring cells towards and around it, whereas at short interparticle distances cells experience steric repulsions. 
\\As a side note, we remark that HIs have also been intensively studied to understand the problem of the coordinated or synchronized motion of beating (or circularly moving) cilia \cite{golestanian2011hydrodynamic, uchida2011generic, friedrich2016hydrodynamic, uchida2017synchronization, maestro2018control, liao2021energetics, solovev2022synchronization}.

\section{Spinners}
We finally discuss the  case where particles self-rotate ($\omega \neq 0$) but do not self-propel ($v_0=0$).  
To become active, such particles require a mechanism coupling rotational and translational degrees of freedom. 
Such a case has been explored in various dry model systems  \cite{tsai2005chiral,van2016spatiotemporal, aragones2016elasticity, workamp2018symmetry,  han2021fluctuating,pietzonka2021oddity}, exhibiting edge currents and crystal melting as a consequence of particle spinning. 
For instance, ref. \cite{han2021fluctuating} considers 2D athermal disks which all spin in the same direction but do not show any individual active translational motion. However, when colliding, the friction between the spinning disks can transform rotational motion into translational one, leading to time-reversal symmetry breaking for the translational degrees of freedom. This is described by the following minimal model for particles of mass $m$ and moment of inertia $I$ moving and spinning within the $(x,y)$-plane
\begin{equation}\label{eq:spin}
m\ddot{\vec{r}}_i =\sum_{j\in \partial_i}\vec{F}_{ij},\,\  
I\dot{\vec{\omega}}_i = \gamma_r(\vec{\omega}-\vec{\omega}_i)+\sum_{j\in \partial_i}\vec{r}_{ij}\times\vec{F}_{ij},
\end{equation}
where the sum runs over particles at contact with the $i$-th particle.
The pairwise forces $\vec{F}_{ij}=-\nabla_i U(r_{ij})-\gamma\vec{v}_{ij}+\gamma\vec{\omega}_{ij}\times \vec{r}_{ij}$ include a conservative part,
deriving from a potential $U$, 
and a non-conservative part, stemming from the chiral torque induced by the spinning of the disks upon collision. Here 
$\vec{\omega}_{ij}$ and $\vec{v}_{ij}$ are the mean angular velocity and the velocity difference  of disks $i,j$, respectively. 
Using particle based simulations and 
coarse-grained  continuum equations, it has been found that (for overdamped rotational dynamics), the system can be (approximately) described in many ways like an equilibrium gas: It 
features a Maxwell-Boltzmann distribution of velocities  with an effective temperature; a Boltzmann distribution of particle densities in external potentials and an ideal gas equation of state. 
However, if perturbed by external forces, unlike for equilibrium systems, the linear response of the system is not characterized by a symmetric viscosity tensor (Onsager relations) which reflects time-reversal symmetry, but features additional off-diagonal terms breaking the even symmetry of the tensor. Such terms give rise to the so-called odd (or Hall) viscosity,
which is a non-dissipative,
transverse stress, which couples shear stresses in different directions.  
Note that it is an exclusive property of 2D systems that such an odd viscosity can occur even in isotropic systems (states) \cite{avron1998odd}.
\\Besides from frictional contacts, mutual torques between spinning particles can originate also from HIs. This has been demonstrated very recently for colloidal magnets actuated by a rotating magnetic field with frequency $\omega$ \cite{soni2019, massana2021arrested, hernandez2022dynamics, bililign2022motile,joshi2022extension} (see Fig. 4). In these experiments the strength of the mutual torques among the colloids decays with distance, which has been explored at the far-field level in refs.~\cite{fily2012cooperative, massana2021arrested}.
\begin{figure}
    \centering
    \includegraphics[width=0.49\textwidth]{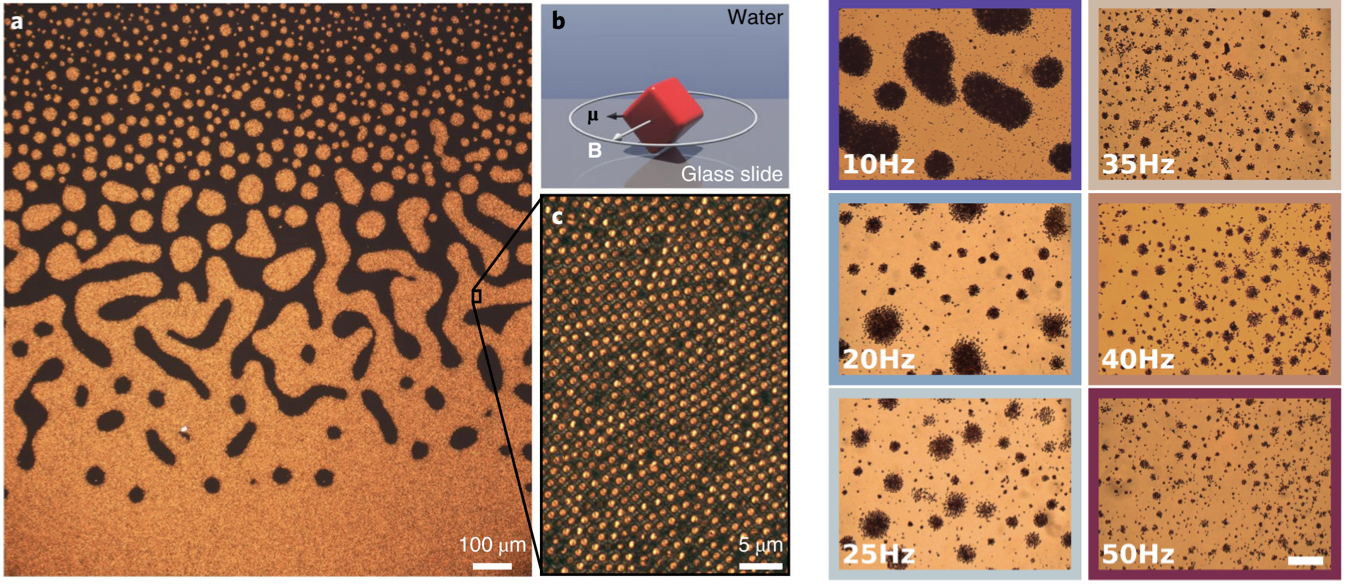}
    \caption{\small Snapshots and illustration of a chiral active fluid of colloidal spinners made of hematite cubes (adapted from \cite{soni2019}) (left three panels) or ellipsoids (adapted from \cite{massana2021arrested}) (right six panels) driven by an external magnetic field, the frequency of which determines the length scale of the emerging structures (right).
    }
\label{fig:soni}
\end{figure}
\\The experiments in ref.~\cite{soni2019} have also been used to test 
hydrodynamic continuum theories of self-spinning particles with odd transport coefficients in the relevant constitutive relations
\cite{avron1998odd,furthauer2012active, furthauer2013active,banerjee2017odd}. The 2D chiral liquid which has been explored in ref. \cite{soni2019},
resembles common Newtonian fluids on several points: it spreads under gravity; nearby droplets merge; voids collapse \cite{soni2019} and it shows phase separation and coarsening \cite{massana2021arrested,joshi2022extension}. However, odd viscosity in the chiral active fluid, causes spontaneous unidirectional flows at the edge of a droplet, as well as 
anomalous attenuation of surface waves \cite{soni2019}. Also, as opposed to canonical equilibrium phase separation, coarsening can be arrested (for fast spinning), leading to finite clusters whose size scales as the inverse of the spinning rate $\ell\propto\omega^{-1}$ \cite{massana2021arrested}. 

\section{Open questions and future challenges}

\emph{Phase diagram and critical behaviour:} 
Despite much progress, 
some important questions regarding the phase diagram of chiral generalizations of key models in active matter (ABP, Vicsek model) 
are still open: What is the nature of the transition between  
micro-flocks (rotating packets), macro-droplets and the vortex phase? 
To which extend can the phase diagram of CAPs be understood based on mean-field theories and linear stability analyses?\footnote{It would be interesting to (numerically) generalize previously performed "instantaneous" linear stability analyses as in ref.~\cite{Liebchen2017Collective} (i) to perform a full Floquet analysis, (ii) to account for perturbations pointing in arbitrary directions and (iii) to test the stability of the different phases which have been observed for CAPs.} 
When are several of these patterns simultaneously stable? (See  ref.~\cite{Liebchen2017Collective} for the decisive role of the initial state.)
Do micro-flock patterns persist, asymptotically for long times, or are they coarsening (unusually slowly) towards macro-droplets? 
What is the impact of chirality on the critical behaviour (and possible universality classes) of active systems?
\\
\emph{Synchronization:}
Previous works suggest that 
activity can induce synchronization \cite{levis2019activity} but also show that in sufficiently large systems the homogeneous synchronized state tends to be linearly unstable \cite{ventejou2021susceptibility}. This raises the question under which conditions (system size, density) 
activity can induce synchronization both for the homogeneous base states (mutual flocking state) and for rotating macro-droplets. 
\\
\emph{Generic scaling relations:} 
As we have seen CAPs with different types of interactions (isotropic, polar) can self-organize into structures with a characteristic length scale $\ell \propto v_0/\omega$. In addition, the scaling 
 $\ell\propto\omega^{-1}$ has been observed even in spinners. Is there a universal mechanism explaining this generic scaling relation? 
\\
\emph{Spinner-limit:}
How does the behavior of a system of aligning circle (helical) swimmers change when approaching
the dynamics of spinners (linear swimmers), i.e. when the average radius goes to zero, {\it either} via $v_0\to 0$ {\it or} via $\omega\to \infty$?
\\
\emph{Large-scale simulations with HIs:} The present literature is surprisingly sparse regarding large scale numerical simulations of wet CAPs with HIs,  which could create a bridge between dry minimal models of CAPs and experiments which have been carried out in suspensions of sperm cells and bacteria.
\\
\emph{Rheology:}
Is there a Hall viscosity in circle (helical) swimmers as in spinner systems?
Which rheological effects can arise from the interplay of chirality and self-propulsion? 
 
\section{Acknowledgments}
D.L. acknowledges MICINN/AEI/FEDER for financial support under grant agreement RTI2018-099032-J-I00.

\bibliographystyle{eplbib}


\begin{thebibliography}{}
\expandafter\ifx\csname url\endcsname\relax\def\url#1{\texttt{#1}}\fi

\end{thebibliography}


\begin{thebibliography}{}
\expandafter\ifx\csname url\endcsname\relax\def\url#1{\texttt{#1}}\fi


\bibitem{bronn1862klassen}\Name{Bronn, H.G.}\REVIEW{Klassen und Ordnungen des Thier-Reichs (CF Winter)}{}{1862}{}
\bibitem{jennings1901significance}\Name{Jennings, H.S.}\REVIEW{The American Naturalist}{35}{1901}{369}
\bibitem{teeffelen2008dynamics}\Name{Van Teeffelen S.\etal}\REVIEW{Phys. Rev. E}{78}{2008}{020101}
\bibitem{Kummel2013Circular}\Name{K{\"u}mmel F.\etal}\REVIEW{Phys. Rev. Lett.}{110}{2013}{198302}
\bibitem{lowen2016chirality}\Name{L{\"o}wen H.}\REVIEW{Eur. Phys. J. Spec. Top.}{225}{2016}{2319}
\bibitem{lei2019nonequilibrium}\Name{Lei Q.\etal}\REVIEW{Sci. Adv.}{5}{2019}{eaau7423}


\bibitem{Liebchen2017Collective}\Name{Liebchen B.\etal}\REVIEW{Phys. Rev. Lett.}{119}{2017}{058002}
\bibitem{levis2019activity}\Name{Levis D.\etal}\REVIEW{Phys. Rev. Res.}{1}{2019}{023026}
\bibitem{levis2019simultaneous}\Name{Levis D.\etal}\REVIEW{Phys. Rev. E}{100}{2019}{012406}
\bibitem{denk2016active}\Name{Denk J.\etal}\REVIEW{Phys. Rev. Lett.}{116}{2016}{178301}
\bibitem{liao2018transport}\Name{Liao, J.-J.\etal}\REVIEW{J. Chem. Phys.}{148}{2018}{094902}
\bibitem{huang2020dynamical}\Name{Huang, Z.-F.\etal}\REVIEW{Phys. Rev. Lett.}{125}{2020}{218002}
\bibitem{kruk2020traveling}\Name{Kruk N.\etal}\REVIEW{Phys. Rev. E}{102}{2020}{022604}
\bibitem{ventejou2021susceptibility}\Name{Ventejou B.\etal}\REVIEW{Phys. Rev. Lett.}{127}{2021}{238001}
\bibitem{Chen2017Weak}\Name{Chong C.\etal}\REVIEW{Nature}{542}{2017}{210}
\bibitem{Kim2018Large}\Name{Kyongwan K.\etal}\REVIEW{Soft Matter}{14}{2018}{3221}
\bibitem{oliver2018synchronization}\Name{Oliver N.\etal}\REVIEW{Soft Matter}{14}{2018}{3073}
\bibitem{Hernandez2020Collective}\Name{Hernández, R.J.\etal}\REVIEW{Soft Matter}{16}{2020}{7704}
\bibitem{Hokmabad2022Spontaneously}\Name{Hokmabad, B.V.\etal}\REVIEW{Soft Matter}{18}{2022}{2731}
\bibitem{afroze2021monopolar}\Name{Afroze F.\etal}\REVIEW{Biochem. Bioph. Res. Co.}{563}{2021}{73}
\bibitem{ma2022dynamical}\Name{Ma Z.\etal}\REVIEW{J. Chem. Phys.}{156}{2022}{021102}
\bibitem{fruchart2021non}\Name{Fruchart M.\etal}\REVIEW{Nature}{592}{2021}{363}
\bibitem{o2017oscillators}\Name{O’Keeffe, K.P.\etal}\REVIEW{Nat. Commun.}{8}{2017}{1}
\bibitem{hoell2017dynamical}\Name{Hoell C.\etal}\REVIEW{New J. Phys.}{19}{2017}{125004}
\bibitem{ai2018mixing}\Name{Ai, B.-Q.\etal}\REVIEW{Soft Matter}{14}{2018}{4388}
\bibitem{maitra2019spontaneous}\Name{Maitra A.\etal}\REVIEW{Nat. Commun.}{10}{2019}{1}
\bibitem{maitra2020chiral}\Name{Maitra A.\etal}\REVIEW{Phys. Rev. Lett.}{125}{2020}{238005}
\bibitem{wang2021emergent}\Name{Wang G.\etal}\REVIEW{Phys. Rev. Lett.}{126}{2021}{108002}
\bibitem{liao2018clustering}\Name{Liao, G.-J.\etal}\REVIEW{Soft Matter}{14}{2018}{7873}
\bibitem{reichhardt2019reversibility}\Name{Reichhardt C.\etal}\REVIEW{J. Chem. Phys.}{150}{2019}{064905}
\bibitem{lowen2019active}\Name{L{\"o}wen H.}\REVIEW{Phys. Rev. E}{99}{2019}{062608}
\bibitem{liu2019configuration}\Name{Liu X.\etal}\REVIEW{J. Chem. Phys.}{151}{2019}{174904}
\bibitem{o2019review}\Name{O'Keeffe K.\etal}\REVIEW{Proc. Spie.}{10982}{2019}{383}
\bibitem{liu2019collective}\Name{Liu Y.\etal}\REVIEW{Soft Matter}{15}{2019}{2999}
\bibitem{kole2021layered}\Name{Kole, S.J.\etal}\REVIEW{Phys. Rev. Lett.}{126}{2021}{248001}
\bibitem{supekar2021learning}\Name{Supekar R.\etal}\REVIEW{arXiv:2101.06568}{}{2021}{}
\bibitem{beppu2021edge}\Name{Beppu K.\etal}\REVIEW{Proc. Natl. Acad. Sci.}{118}{2021}{e2107461118}
\bibitem{moore2021chiral}\Name{Moore, J.M.\etal}\REVIEW{Soft Matter}{17}{2021}{4559}
\bibitem{kruk2021finite}\Name{Kruk N.\etal}\REVIEW{J. Comput. Phys.}{440}{2021}{110275}
\bibitem{zhang2021persistence}\Name{Zhang B.\etal}\REVIEW{Soft Matter}{17}{2021}{4818}
\bibitem{reigh2020active}\Name{Reigh, S.Y.\etal}\REVIEW{Soft Matter}{16}{2020}{1236}
\bibitem{hernandez2022dynamics}\Name{Hern{\'a}ndez, R.J.\etal}\REVIEW{Appl. Phys. Lett.}{120}{2022}{081601}
\bibitem{mijalkov2013sorting}\Name{Mijalkov M.\etal}\REVIEW{Soft Matter}{9}{2013}{6376}
\bibitem{kurzthaler2017intermediate}\Name{Kurzthaler C.\etal}\REVIEW{Soft Matter}{13}{2017}{6396}
\bibitem{chepizhko2020random}\Name{Chepizhko O.\etal}\REVIEW{New J. Phys.}{22}{2020}{073022}
\bibitem{heckel2020}\Name{Heckel S.\etal}\REVIEW{Langmuir}{36}{2020}{12473}
\bibitem{scholz2021surfactants}\Name{Scholz, C. \etal}\REVIEW{Sci. Adv.}{7}{2021}{eabf8998}
\bibitem{markovich2019chiral}\Name{Markovich, T. \etal}\REVIEW{New J. Phys.}{21}{2019}{112001}

\bibitem{marchetti2013hydrodynamics}\Name{Marchetti, M.C.\etal}\REVIEW{Rev. Mod. Phys.}{85}{2013}{1143}
\bibitem{bechinger2016active}\Name{Bechinger C.\etal}\REVIEW{Rev. Mod. Phys.}{88}{2016}{045006}
\bibitem{chate2020dry}\Name{Chat{\'e} H.}\REVIEW{Annu. Rev. Conden. Ma. P.}{11}{2020}{189}
\bibitem{hecht2021introduction}\Name{Hecht L.\etal}\REVIEW{arXiv:2102.13007}{}{2021}{}
\bibitem{liebchen2021interactions}\Name{Liebchen B.\etal}\REVIEW{J. Phys. Condens. Matter}{34}{2021}{083002}
\bibitem{riedel2005self}\Name{Riedel, I.H.\etal}\REVIEW{Science}{309}{2005}{300}
\bibitem{friedrich2007chemotaxis}\Name{Friedrich, B.M.\etal}\REVIEW{Proc. Natl. Acad. Sci.}{104}{2007}{13256}
\bibitem{patra2022collective}
\Name{Patra P.\etal}\REVIEW{Nat. Phys.}{18}{2022}{586}
\bibitem{diluzio2005escherichia}\Name{DiLuzio, W.R.\etal}\REVIEW{Nature}{435}{2005}{1271}
\bibitem{lauga2006swimming}\Name{Lauga E.\etal}\REVIEW{Biophys. J.}{90}{2006}{400}
\bibitem{Leonardo2011Swimming}\Name{Di Leonardo R.\etal}\REVIEW{Phys. Rev. Lett.}{106}{2011}{038101}
\bibitem{ten2014gravitaxis}\Name{Ten, H.B.\etal}\REVIEW{Nat. Commun.}{5}{2014}{1}
\bibitem{Shelke2019Transition}\Name{Yogesh S.\etal}\REVIEW{Langmuir}{35}{2019}{4718}
\bibitem{Zhang2020Reconf}\Name{Zhang B.\etal}\REVIEW{Nat. Commun.}{11}{2020}{1}
\bibitem{alvarez2021reconfigurable}\Name{Alvarez L.\etal}\REVIEW{Nat. Commun.}{12}{2021}{1}
\bibitem{barois2020sorting}\Name{Barois T.\etal}\REVIEW{Phys. Rev. Lett.}{125}{2020}{238003}
\bibitem{arora2021emergent}\Name{Arora P.\etal}\REVIEW{Sci. Adv.}{7}{2021}{eabd0331}
\bibitem{Kruger2016Curling}\Name{Krüger C.\etal}\REVIEW{Phys. Rev. Lett.}{117}{2016}{048003}
\bibitem{Lancia2019Reorientation}\Name{Lancia F.\etal}\REVIEW{Nat. Commun.}{10}{2019}{1}
\bibitem{Wang2021Active}\Name{Xin W.\etal}\REVIEW{Soft Matter}{17}{2021}{2985}
\bibitem{van2016spatiotemporal}\Name{van Zuiden, B.C.\etal}\REVIEW{Proc. Natl. Acad. Sci.}{113}{2016}{12919}
\bibitem{scholz2018rotating}\Name{Scholz C.\etal}\REVIEW{Nat. Commun.}{9}{2018}{1}
\bibitem{workamp2018symmetry}\Name{Workamp M.\etal}\REVIEW{Soft Matter}{14}{2018}{5572}
\bibitem{soni2019}\Name{Soni V.\etal}\REVIEW{Nat. Phys.}{15}{2019}{1188}
\bibitem{massana2021arrested}\Name{Massana-Cid H.\etal}\REVIEW{Phys. Rev. Res.}{3}{2021}{L042021}
\bibitem{bililign2022motile}\Name{Bililign, E.S.\etal}\REVIEW{Nat. Phys.}{18}{2022}{212}
\bibitem{kraft2013brownian}\Name{Kraft, D.J.\etal}\REVIEW{Phys. Rev. E}{88}{2013}{050301}
\bibitem{ebbens2010self}\Name{Ebbens S.\etal}\REVIEW{Phys. Rev. E}{82}{2010}{015304}
\bibitem{nourhani2016spiral}\Name{Nourhani A.\etal}\REVIEW{Phys. Rev. E}{94}{2016}{030601}
\bibitem{wykes2016dynamic}\Name{Wykes, M.S.D.\etal}\REVIEW{Soft Matter}{12}{2016}{4584}
\bibitem{vutukuri2017rational}\Name{Vutukuri, H.R.\etal}\REVIEW{Sci. Rep. Uk.}{7}{2017}{1}
\bibitem{aubret2018targeted}\Name{Aubret A.\etal}\REVIEW{Nature Phys.}{14}{2018}{1114}
\bibitem{aubret2018diffusiophoretic}\Name{Aubret A.\etal}\REVIEW{Soft Matter}{14}{2018}{9577}
\bibitem{schmidt2019light}\Name{Schmidt F.\etal}\REVIEW{J. Chem. Phys.}{150}{20
19}{094905}
\bibitem{grauer2021active}\Name{Grauer J.\etal}\REVIEW{Nature Commun.}{12}{2021}{1}
\bibitem{narinder2018memory}\Name{Narinder N.\etal}\REVIEW{Phys. Rev. Lett.}{121}{2018}{078003}
\bibitem{weeks1971role}\Name{Weeks J.D. \etal}\REVIEW{J. Chem. Phys.}{54}{1971}{5237}
\bibitem{wittkowski2012self}\Name{Wittkowski R.\etal}\REVIEW{Phys. Rev. E}{85}{2012}{021406}
\bibitem{CatesTailleurRev}\Name{Cates, M.E.\etal}\REVIEW{Annu. Rev. Condens. Matter Phys.}{6}{2015}{219}
\bibitem{mani2022}\Name{Ma Z.\etal}\REVIEW{J. Chem. Phys.}{156}{2021}{021102}
\bibitem{bickmann2020analytical}\Name{Bickmann J.\etal}\REVIEW{arXiv:2010.05262}{}{2020}{}
\bibitem{ElenaJCP}\Name{Sese-Sansa E.\etal}\REVIEW{arXiv:2204.11571}{}{2022}{}
\bibitem{schwarz2012phase}\Name{Schwarz-Linek J.\etal}\REVIEW{Proc. Natl. Acad. Sci.}{109}{2012}{4052}
\bibitem{BialkeEPL}\Name{Bialk{\'{e}} J.\etal}\REVIEW{EPL}{103}{2013}{1}
\bibitem{Vicsek1995}\Name{Vicsek T.\etal}\REVIEW{Phys. Rev. Lett.}{75}{1995}{1226}
\bibitem{farrell2012pattern}\Name{Farrell, F.D.C.\etal}\REVIEW{Phys. Rev. Lett.}{108}{2012}{248101}
\bibitem{martin2018collective}\Name{Mart{\'\i}n-G{\'o}mez A.\etal}\REVIEW{Soft Matter}{14}{2018}{2610}
\bibitem{chepizhko2021revisiting}\Name{Chepizhko O.\etal}\REVIEW{Soft Matter}{17}{2021}{3113}
\bibitem{loose2014bacterial}\Name{Loose M.\etal}\REVIEW{Nat. Cell Biol.}{16}{2014}{38}
\bibitem{dean1996langevin}\Name{Dean, D.S.}\REVIEW{J. Phys. A}{29}{1996}{L613}
\bibitem{risken1996fokker}\Name{Risken, H.}\REVIEW{The Fokker-Planck-Equation (Springer)}{}{1996}{}
\bibitem{bertin2006boltzmann}\Name{Bertin E. \etal}\REVIEW{Phys. Rev. E}{74}{2006}{022101}
\bibitem{peshkov2014}\Name{Peshkov, A. \etal}\REVIEW{Eur. Phys. J. Spec. Top.}{223}{2014}{1315}
\bibitem{liao2021emergent}\Name{Liao, G.-J.\etal}\REVIEW{Soft Matter}{17}{2021}{6833}
\bibitem{levis2018micro}\Name{Levis D.\etal}\REVIEW{J. Phys. Condens. Mat.}{30}{2018}{084001}
\bibitem{liebchen2016pattern}\Name{Liebchen B.\etal}\REVIEW{Soft Matter}{12}{2016}{7259}
\bibitem{blake1971spherical}\Name{Blake, J.R.}\REVIEW{J. Fluid. Mech.}{46}{1971}{199}
\bibitem{pedley2016squirmers}\Name{Pedley, T.J.\etal}\REVIEW{J. Fluid. Mech.}{798}{2016}{165}
\bibitem{paklauga}\Name{Pak, O.S.\etal}\REVIEW{J. Eng. Math.}{88}{2014}{1}
\bibitem{burada2022hydrodynamics}\Name{Burada, P.S.\etal}\REVIEW{Phys. Rev. E}{105}{2022}{024603}
\bibitem{maity2022near}\Name{Maity R.\etal}\REVIEW{arXiv:2204.07512}{}{2022}{}
\bibitem{lauga2019transition}\Name{Das D.\etal}\REVIEW{Phys. Rev. E}{100}{2019}{043117}
\bibitem{elgeti2010hydrodynamics}\Name{Elgeti J.\etal}\REVIEW{Biophys. J.}{99}{2010}{1018}
\bibitem{drescher2009dancing}\Name{Drescher K.\etal}\REVIEW{Phys. Rev. Lett.}{102}{2009}{168101}
\bibitem{shen2019hydrodynamic}\Name{Shen Z.\etal}\REVIEW{Soft matter}{15}{2019}{1508}
\bibitem{petroff2015fast}\Name{Petroff, A.P.\etal}\REVIEW{Phys. Rev. Lett.}{114}{2015}{158102}
\bibitem{golestanian2011hydrodynamic}\Name{Golestanian R.\etal}\REVIEW{Soft Matter}{7}{2011}{3074}
\bibitem{uchida2011generic}\Name{Uchida N.\etal}\REVIEW{Phys. Rev. Lett.}{106}{2011}{058104}
\bibitem{maestro2018control}\Name{Maestro A.\etal}\REVIEW{Commun. Phys.}{1}{2018}{1}
\bibitem{solovev2022synchronization}\Name{Solovev A.\etal}\REVIEW{Chaos}{32}{2022}{013124}
\bibitem{uchida2017synchronization}\Name{Uchida N.\etal}\REVIEW{J. Phys. Soc. Jpn.}{86}{2017}{101007}
\bibitem{friedrich2016hydrodynamic}\Name{Friedrich B.}\REVIEW{Eur. Phys. J. Spec. Top.}{225}{2016}{2353}
\bibitem{liao2021energetics}\Name{Liao W.\etal}\REVIEW{Phys. Rev. E}{103}{2021}{042419}
\bibitem{tsai2005chiral}\Name{Tsai, J.-C.\etal}\REVIEW{Phys. Rev. Lett.}{94}{2005}{214301}
\bibitem{aragones2016elasticity}\Name{Aragones, J.L.\etal}\REVIEW{Nat. Commun.}{7}{2016}{1}
\bibitem{han2021fluctuating}\Name{Han M.\etal}\REVIEW{Nat. Phys.}{17}{2021}{1260}
\bibitem{pietzonka2021oddity}\Name{Pietzonka P.}\REVIEW{Nat. Phys.}{17}{2021}{1193}
\bibitem{avron1998odd}\Name{Avron, J.E.}\REVIEW{J. Stat. Phys.}{92}{1998}{543}
\bibitem{joshi2022extension}\Name{Joshi K.\etal}\REVIEW{Proc. Natl. Acad. Sci.}{119}{2022}{e2117971119}
\bibitem{fily2012cooperative}\Name{Fily Y.\etal}\REVIEW{Soft Matter}{8}{2012}{3002}
\bibitem{furthauer2012active}\Name{F{\"u}rthauer S.\etal}\REVIEW{Eur. Phys. J. E}{35}{2012}{1}
\bibitem{furthauer2013active}\Name{F{\"u}rthauer S.\etal}\REVIEW{Phys. Rev. Lett.}{110}{2013}{048103}
\bibitem{banerjee2017odd}\Name{Banerjee D.\etal}\REVIEW{Nat. Commun.}{8}{2017}{1}








\end{thebibliography}

\end{document}